\documentstyle[prl,epsfig,multicol,aps]{revtex}

\begin{document}
\draft

\title{Strong fragmentation of low-energy electromagnetic excitation 
strength in $^{117}$Sn}
\author{
V. Yu. Ponomarev$^{1}$ \cite{byline1},
J. Bryssinck$^{1}$, L. Govor$^{2}$, 
F. Bauwens$^{1}$, O. Beck$^{3}$, D. Belic$^{3}$, P. von
Brentano$^{4}$, 
D. De Frenne$^{1}$, \\
C. Fransen$^{4}$, R.-D. Herzberg$^{4,5}$,
E. Jacobs$^{1}$, U. Kneissl$^{3}$, H. Maser$^{3}$, A. Nord$^{3}$, 
N. Pietralla$^{4}$, H.H. Pitz$^{3}$, V. Werner$^{4}$}
\address{$^{1}$Vakgroep Subatomaire en Stralingsfysica, 
Universiteit Gent,
Proeftuinstraat 86,
9000 Gent, Belgium}
\address{$^{2}$Russian Scientific Centre ``Kurchatov Institute'',
Moscow, Russia}
\address{$^{3}$Institut f\"ur Strahlenphysik, Universit\"{a}t Stuttgart,
Stuttgart, Germany}
\address{$^{4}$ Institut f\"ur Kernphysik, Universit\"{a}t zu K\"{o}ln,
K\"{o}ln, Germany.}
\address{$^{5}$Oliver Lodge Laboratory, University of Liverpool,
Oxford Street, Liverpool, L69 7ZE, UK}
\date{\today }
\maketitle

\begin{abstract}
Results of nuclear resonance fluorescence experiments on
$^{117}$Sn are reported. 
More than 50 $\gamma$ transitions with $E_{\gamma} < 4$~MeV were detected  
indicating a strong fragmentation of the electromagnetic excitation strength.
For the first time microscopic
calculations  making use of a complete
configuration space for low-lying states are performed 
in heavy odd-mass spherical nuclei. 
The theoretical predictions are in good agreement with the 
 data.
It is concluded that although the $E1$ transitions are the strongest ones
also $M1$ and $E2$ decays contribute substantially to the observed spectra.
In contrast to the neighboring even $^{116-124}$Sn, in $^{117}$Sn
the $1^-$ component of
the two-phonon $[2^+_1 \otimes 3^-_1]$ quintuplet built on top of the 
1/2$^+$ ground state is proved to be strongly fragmented.
\end{abstract}


\begin{multicols}{2}
\narrowtext

During the last years the
study of the properties of multi-phonon excitations in atomic nuclei 
has been one of the key topics in nuclear structure research. 
The main motivation for these studies is to gain insight into the
applicability of a picture of harmonic excitations to 
a finite strongly interacting fermion system, like an atomic nucleus. 
At present, systematic experimental information on the properties 
of low-lying two-phonon states \cite{Kne96,Her95,Bry99} as well as 
on two-phonon giant resonances \cite{Aum98,Ber99} in even-even 
spherical nuclei is available.

The $1^-$ component of a 
two-phonon quintuplet built up of collective quadrupole and octupole
phonons, $[2^+_1 \otimes 3^-_1]$, is a very good candidate for the
investigation of anharmonic properties of low-lying excited states.
This state is practically a pure two-phonon
state with very weak admixtures of other configurations \cite{Bry99}.
In addition, a tool which is very selective to its excitation, 
namely nuclear resonance fluorescence (NRF), is available.
The main conclusion from systematic studies 
\cite{Her95,Bry99,Wil98,End98,Pie99}
of this $1^-$ state in even
spherical nuclei in different mass regions is that anharmonicity 
effects are rather weak, supporting a harmonic treatment of nuclear 
excitations.

The next step in these studies should be the extension to spherical odd-mass
nuclei. Here experimental information is sparse. 
NRF experiments on $^{113}$Cd \cite{Cd}, $^{133}$Cs \cite{Cs}
and $^{143}$Nd \cite{Nd}
have been reported. From very general arguments one expects  
rather weak changes in the properties of collective vibrations when 
they are coupled to an odd quasiparticle. 
However, even the first NRF experiments 
indicate a rather strong fragmentation of the excitation strength. 
To minimize the number of possible excited levels it is imperative
to consider this coupling of collective vibrations to an odd
quasiparticle in a nucleus with ground state spin 1/2.

In the present letter we report on NRF experiments on $^{117}$Sn 
($J^{\pi}_{g.s.} =1/2^+$) 
in which 59 transitions with $E_{\gamma} < 4.05$~MeV 
have been observed.
It is clear that 
a detailed theoretical analysis which provides information on the
structure of the excited states,  is necessary to explain the
results of this experiment. 
For this reason a full microscopic calculation
of the properties of multi-phonon states in heavy odd-mass spherical nuclei
has been performed for the first time. 

Experimental results have been obtained at the bremsstrahlung NRF-facility
at the 4.3 MV Dynamitron accelerator of the Stuttgart University 
\cite{Kne96}. The electron energy was 4.1 MeV. 
The NRF technique has already been extensively described in review articles
(see, e.g. Ref.~\cite{Kne96}).
For this experiment, a setup with 3 HP Ge-detectors with 100\%
efficiency (relative to a 3''$\times$3'' Na(Tl) crystal) and installed 
at  scattering angles 90$^{\circ}$, 127$^{\circ}$ and 150$^{\circ}$
was used. The NRF target consisted of two tin disks. 
The total amount of tin was 1.649 g and the enrichment in $^{117}$Sn 
was 92.10\%. Two $^{27}$Al disks with a total weight of
780 mg were sandwiched between the tin disks. 
The aluminum nuclei, having well known photon scattering cross sections
\cite{Pie95}, enable the photon flux calibration and thereby 
the absolute measurement of the photon scattering cross sections of
excited states in the $^{117}$Sn nuclei.

\begin{figure}[tbhp]
\begin{center}
\epsfig{figure=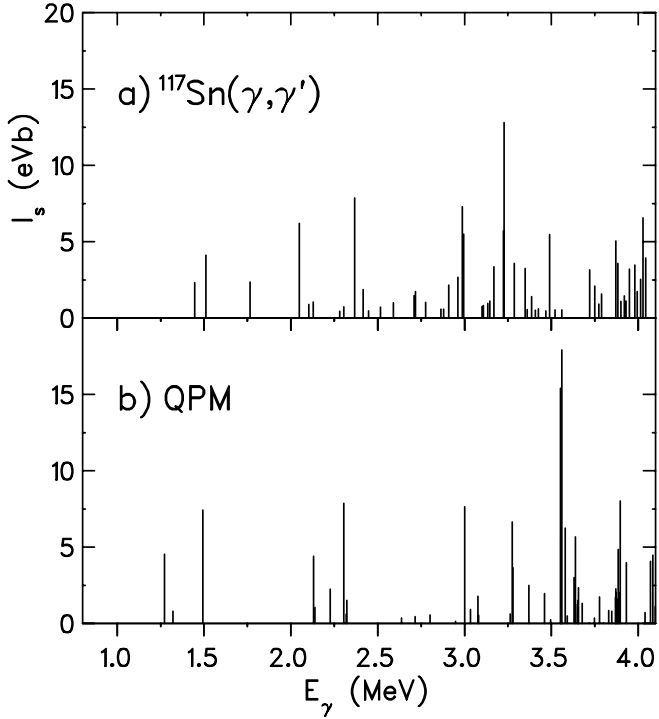,width=8.6cm,angle=0}
\end{center}
\caption{a) Experimental and b) calculated intensities of 
$\gamma$-decays of low-lying states in $^{117}$Sn
into the ground state. 
\label{fig1} }
\end{figure}

In this experiment 59 $\gamma$-transitions with an energy between 1440 and 
4050~keV have been detected. From the consideration of $\gamma$ energy 
differences it follows that only 5 of the observed $\gamma$
transitions may be due to inelastic photon scattering.
The energy integrated cross sections for elastic scattering, $I_s$,
for all observed states are plotted in Fig.~\ref{fig1}a.
Spin quantum numbers could be assigned to some excited states from 
angular distribution measurements:
for 6 levels the spin
value 3/2 is suggested and for 
another 5 levels an assignment 1/2(3/2), meaning that 1/2 is more
likely, is possible.
The present NRF setup does not allow to deduce parities for levels in 
odd-mass nuclei.  
In the energy region under consideration
only a few levels are known from other experiments \cite{117Sn}.
More details on the experimental analysis and the complete information 
on observed transitions will be published in a forthcoming 
article.

Combining the results of the present experimental studies with 
those for $^{113}$Cd \cite{Cd}, $^{133}$Cs \cite{Cs} 
and $^{143}$Nd \cite{Nd} we
conclude that a strong fragmentation of low-energy electromagnetic
excitation strength is a general feature  
in heavy odd-mass nuclei.
With the present state-of-the-art of the modern NRF technique alone 
it is not possible to study in detail the properties of the many levels
involved and thus, a theoretical support is needed.   

To accomplish this task
a theoretical analysis of the properties of low-lying states in $^{117}$Sn 
has been performed within the Quasiparticle-Phonon Model (QPM)
\cite{Sol}.
The QPM has been successfully applied to describe the position and 
the $E1$ excitation probability of the lowest $1^-$ state in the 
even-even $^{116-124}$Sn isotopes \cite{Bry99}.
This state has a two-phonon nature with a 
contribution of the $[2^+_1 \otimes 3^-_1]_{1^-}$ configuration of
96-99\%. 
A general QPM formalism to treat odd-mass spherical nuclei is presented 
in review articles \cite{Vdo85,Gal88}.
A Woods-Saxon potential is used in QPM as an average field for
protons and neutrons. Phonons of different multipolarities and
parities are obtained by solving RPA equations with a separable form
of the residual interaction. The
single-particle spectrum and phonon basis are fixed from  calculations 
in the neighboring even-even nuclear core. No new free parameters are
introduced for calculations in the odd-mass nuclei. All 
matrix elements of the coupling
between the different configurations in the wave functions of 
states in odd-mass nuclei are calculated on a microscopic footing, making
use of the internal fermion structure of the phonons and the model 
Hamiltonian.
In the calculations to be presented below we use the same set of
parameters as in our previous calculations in $^{116}$Sn \cite{Bry99}.

In our present calculations the wave functions of the ground state and 
excited states are mixtures of different 
``quasiparticle $\otimes N$-phonon'' ($[qp \otimes N ph]$)
configurations, where $N=0,1,2,3$. 
When $N \ge 1$, $qp$ from different levels of the average field are
accounted for. It is only necessary that all configurations have the
same total spin and parity. To achieve a correct position 
of the $[qp \otimes 2 ph]$ configurations, in which 
we are especially interested in these studies,
$[qp \otimes 3 ph]$ configurations are important.
The excitation energies and the contribution of different components from
the configuration space to the structure of each excited state are
obtained by a diagonalization of the model Hamiltonian on a set of
employed wave functions. 
Interaction matrix elements are calculated to first order 
perturbation theory. This
means that any $[qp \otimes N ph]$ configuration interacts with the
$[qp \otimes (N \pm 1) ph]$) ones, but its coupling to 
$[qp \otimes (N \pm 2) ph]$ configurations 
is not included in this theoretical treatment.
The omitted couplings have non-vanishing  interaction  matrix elements 
only in second order perturbation theory. They are much smaller
than the ones accounted for and excluded from consideration for
technical reasons. An interaction with other $[qp \otimes N ph]$ 
configurations is taken into account while treating Pauli principle 
corrections. 

The phonon basis includes the phonons with multipolarity and parity
$\lambda^{\pi} =~1^{\pm},~2^+,~3^-$ and $4^+$. Several low-energy phonons 
of each multipolarity are included in the model space. 
The most important ones are the first collective $2^+,~3^-$ 
and $4^+$ phonons 
and the ones which form the giant dipole resonance (GDR).
Non-collective low-lying phonons of an unnatural parity and natural parity
phonons of higher multipolarities are of a marginal importance.
To make realistic calculations possible one has to truncate the
configuration space. We have done this on the basis of excitation energy 
arguments. All $[qp \otimes 1 ph]$ and 
$[qp \otimes 2 ph]$ with $E_x \le 6$~MeV, and 
$[qp \otimes 3 ph]$ with $E_x \le 8$~MeV configurations 
which do not violate the Pauli principle are included in the model space.
The only exception are $[J_{g.s.} \otimes 1^-]$ configurations which
have not been truncated at all to treat a core polarization effect due 
to the coupling of low-energy dipole transitions to the GDR on a microscopic
level. Thus, for electric dipole transitions we have not renormalized 
effective charges and used $e^{{\rm eff}}(p)=(N/A)\,e$ and 
$e^{{\rm eff}}(n)=-(Z/A)\,e$ values to separate the center of mass
motion. 
For $M1$ transitions we use $g_s^{{\rm eff}}= 0.64g_s^{{ \rm free}}$
as recommended in Ref.~\cite{vnc99}.
In this  all the
configurations of importance for the description of low-lying states up to
4~MeV are included in the model space. 
The dimension of this space depends on the
total spin of the excited states, and it varies between 500 and 700.

\begin{figure}[tbhp]
\begin{center}
\epsfig{figure=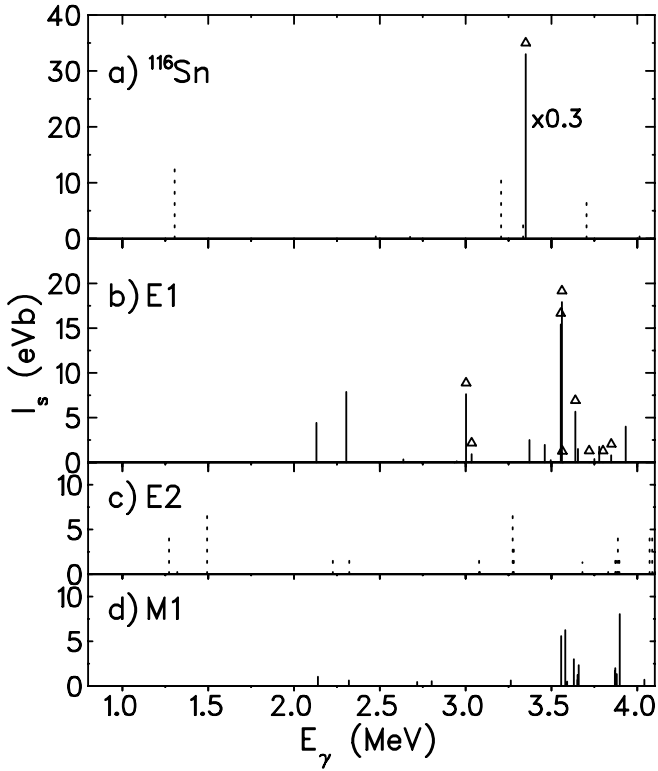,width=8.6cm,angle=0}
\end{center}
\caption{Calculated intensities of $\gamma$-decays into the ground
state in a) $^{116}$Sn and b-d) $^{117}$Sn. 
$E2$ decays are plotted by dashed lines in a) and c).
$E1$ decays in b) which are predominantly due to
$[3s_{1/2} \otimes [2^+_1 \otimes 3^-_1]_{1^-}]_{1/2^-,3/2^-} 
\to 3s_{1/2}$ transitions
are marked by triangles.
\label{fig2} }
\end{figure}

Since only $E1$, $M1$ and $E2$ transitions
can be observed in the present experiment only the properties of
excited states with
$J^{\pi} = 1/2^{\pm},~3/2^{\pm}$ and $5/2^+$ will be presented. 
As the parities of the decaying levels are unknown 
and the total spin of only
a few of them is assigned, the best comparison between theoretical
predictions and experimental data should be at the level of the
scattering cross sections, $I_s$.
The calculated $I_s$ values:
\begin{equation} 
I_s(\lambda) \propto  
\cdot E_x^{2 \lambda - 1} 
\cdot B( \lambda) 
\cdot \frac{\Gamma_{g.s.}}{\Gamma_{tot}}~,
\label{eq}
\end{equation} 
where $\Gamma$ are decay widths,
are plotted in Fig.~\ref{fig1}b. 
We present here only ground state decays. 
The inelastic decays are accounted for in a calculation of the total 
decay width $\Gamma_{tot}$. 
Details concerning the  
calculations and branching ratios will be presented elsewhere.
Supporting the experimental findings our calculations also provide a
strong fragmentation of electromagnetic strength. 
The strongest transitions have $E1$ character, but also $E2$ and $M1$
excitations yield comparable cross sections.
They are presented separately in Fig.~\ref{fig2}b,c,d, respectively, 
in comparison to calculated intensities in the core nucleus, $^{116}$Sn
(Fig.~\ref{fig2}a).
The total calculated scattering  cross sections, $I_s$, of
the plotted $E1$, $M1$ and $E2$ excitations in Fig~\ref{fig2}b-d
equal 73, 37 and 42 eVb.
The summed cross section of the experimentally observed levels,
presented in Fig.1a, equals 137 eVb and agrees within 10\% with
the theoretically predicted 152 eVb.

The  good correspondence between experimental and theoretical
results concerning strength fragmentation, position of main groups of levels
and total electromagnetic strength make us confident in the theoretical
predictions, which will be used below in explaining the main features 
of the experimentally observed transitions.

\begin{figure}[tbhp]
\begin{center}
\epsfig{figure=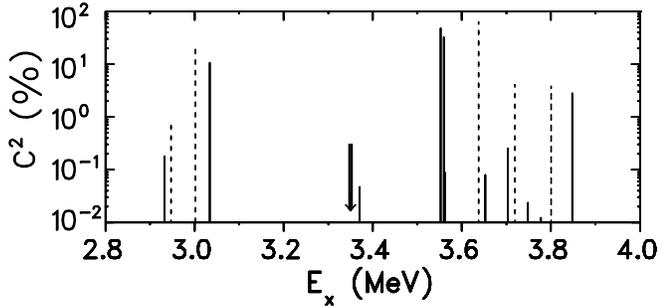,width=8.6cm,angle=0}
\end{center}
\caption{Contribution, $C^2$, of the 
$[3s_{1/2} \otimes [2^+_1 \otimes 3^-_1]_{1^-}]_{J^{\pi}}$ configuration
to wave functions of $J^{\pi} = 3/2^-$ (solid lines) and 
$J^{\pi} = 1/2^-$ (dashed lines) states in $^{117}$Sn. 
The position of the  
$[2^+_1 \otimes 3^-_1]_{1^-}$ state in $^{116}$Sn  is indicated by arrow.
\label{fig3} }
\end{figure}

Let us first ask whether the unpaired quasiparticle in odd-mass
nuclei plays the role of a spectator in determining the properties of
low-lying states. To answer this question we compare the structure of
the excited states in $^{117}$Sn with the structure of the corresponding
states in the neighboring even-even core, $^{116}$Sn. 
We observe the most essential
changes in the states connected to the ground state by $E1$ 
transitions. In the even-even core there is only one $1^-$ configuration
with an excitation energy below 4~MeV (line with triangle in 
Fig.~\ref{fig2}a).
It has a 
$[2^+_1 \otimes 3^-_1]_{1^-}$ two-phonon nature \cite{Bry99}. 
This is a general feature in heavy semi-magic even-even nuclei
\cite{Pon98}.
All other $1^-$ configurations are more
than 1~MeV above. As a result the $1^-_1$ state has practically pure
two-phonon character in semi-magic nuclei. In contrast, 
the two corresponding configurations in $^{117}$Sn
$[3s_{1/2} \otimes [2^+_1 \otimes 3^-_1]_{1^-}]_{1/2^-,3/2^-}$ are
embedded in other $[qp \otimes 1 ph]$ and $[qp \otimes 2 ph]$
configurations with the same spin and parity.
An interaction with them leads to a strong fragmentation 
of these two main configurations (see, Fig.~\ref{fig3}). 
All the resulting states are
carrying a fraction of the $E1$ excitation strength from the ground state.
The main part of the 
$[3s_{1/2} \otimes [2^+_1 \otimes 3^-_1]_{1^-}]_{1/2^-,3/2^-}$ 
configurations is concentrated in 3/2$^-$ states with
an excitation energy of 3.04~MeV (11\%), 3.55~MeV (48\%) and 3.56~MeV
(32\%) and in 1/2$^-$ states at 3.00~MeV (22\%) and
3.63~MeV (63\%) (see, lines with triangles in 
Fig.~\ref{fig2}b). 
The $E1$ strength distribution among low-lying levels is even more
complex because 3/2$^-$ states at 2.13, 2.33 and 3.93~MeV have a
noticeable contribution of the $3p_{3/2}$ one-quasiparticle configuration 
with a reduced excitation matrix element  
$< 3p_{3/2} ||E1|| 3s_{1/2}>$ 
for which there is no analogue  in the even-even core $^{116}$Sn.
Also the coupling to $[3s_{1/2} \otimes 1^-_{\mbox{\tiny GDR}}]$, for
treating the core polarization effect, is somewhat different 
than in the core nucleus, because of a blocking role of the interaction with
other configurations (see, also Ref.~\cite{vnc95}, where only the last
type of transitions have been accounted for in the calculations of the $E1$ 
strength distribution in $^{115}$In). 
The total $B(E1)\uparrow$ strength in the energy region from 2.0 to
4.0~MeV in this calculation equals $7.2 \cdot 10^{-3}$~e$^2$fm$^2$.
It should be compared to the calculated $B(E1, 0^+_{g.s.} \rightarrow 
[2^+ \otimes 3^-]_{1^-}) = 8.2 \cdot 10^{-3}$~e$^2$fm$^2$ in the
neighboring $^{116}$Sn nucleus \cite{Bry99a}.  

Positive parity states in $^{117}$Sn are deexciting to the 1/2$^+$ ground
state by $E2$ and $M1$ transitions. The $B(E2)\uparrow$ strength 
distribution is dominated by the excitation of the 3/2$^+$ state at 1.27~MeV 
and the 5/2$^+$ state at 1.49~MeV.  
The wave functions of these states carry respectively 85\% and 
60\% of the $[3s_{1/2} \otimes 2^+_1]$ configuration.
A smaller
fraction of this configuration can be found
in the  3/2$^+$ state at 2.32~MeV (5\%) and the
5/2$^+$ state at 2.23~MeV (6\%).
The rather fragmented $E2$ strength at higher energies 
(Fig.~\ref{fig2}c) is mainly due to 
$[3s_{1/2} \otimes 2^+_{4,5}]$ configurations which are much less
collective than the first one. Fragmented $E2$ strength between 2.0 and 
4.0~MeV originating from excitation of $2^+_{4,5}$ phonons has
been also observed in NRF experiments on the even $^{116}$Sn 
nucleus \cite{Bry99a}. It is supported by theoretical calculations
(see, dashed lines in Fig.~\ref{fig2}a).
In the odd-mass $^{117}$Sn nucleus it is even more fragmented because of 
the higher density of configurations. 
Nevertheless, these $E2$ excitations at high energies contribute 
appreciably to the reaction cross section, because the $E2$ 
photon scattering cross section is a cubic function 
on the excitation energy (see, Eq.~(\ref{eq})).

The $B(M1)\uparrow$ strength in these calculations is concentrated
mainly above 3.5~MeV as can be seen in Fig.~\ref{fig2}d. 
The wave functions of all the 1/2$^+$ and 3/2$^+$
states at these energies are very complex. The main configurations, 
responsible for the $M1$ strength, are the $[2d_{5/2,3/2} \otimes 2^+_i]$
ones which are excited because of the internal fermion structure of the
phonons (similar to $E1$ 
 $0^+_{g.s.} \to [2^+_1 \otimes 3^-_1]_{1^-}$ excitations).
They have no analogous transitions in even-even nuclei.
The configuration $[3s_{1/2} \otimes 1^+_1]$ has an excitation energy 
of about 4.2~MeV but its contribution to the structure of states below 
4~MeV is rather weak. Most of the states with the largest
$B(M1)\uparrow$ values have $J^{\pi} = 1/2^+$.

To conclude, a strong fragmentation of low-energy electromagnetic
excitation strength in $^{117}$Sn has been observed in NRF experiments.
This observation is understood by microscopic calculations using
a complete configurational space for low-lying states.
The calculations show that several $E1$, $M1$ and $E2$ excitations
lead to comparably large electromagnetic cross sections, which
are expected to be observed in this highly sensitive NRF experiment.
The structure of decaying states is analyzed, showing that the
$[1/2^+_{\rm g.s.}\otimes [2^+\otimes 3^- ]_{1^-}]$
configuration is strongly fragmented.
The responsible configuration mixing is calculated to be
considerably more complex, than it was suggested previously \cite{Nd}
in order to explain the corresponding fragmentation of
electromagnetic excitation strength in the semi-magic
nucleus $^{143}$Nd.

We thank Prof. K.Heyde for fruitful suggestions to this manuscript.
This work is part of the Research program of
the Fund for Scientific Research-Flanders.  The support by the
Deutsche Forschungsgemeinschaft (DFG) under contracts Kn 154-30 and Br
799/9-1 is gratefully acknowledged.
V.Yu. P. acknowledges support from the Research Council 
of the University of Gent.

\end{multicols}
\end{document}